\documentstyle[12pt,aaspp4,astro]{article}

\newcommand{\unit}[1]{\ifmmode\,{\rm #1}\else$\,{\rm #1}$\fi}
\newcommand{\wmsh}{\unit{Wm^{-2}ster^{-1}Hz^{-1}}}
\newcommand{\mum}{\unit{\mu m}}
\newcommand{\mujy}{\unit{\mu Jy}}
\newcommand{\cm}{\unit{cm}}
\newcommand{\etal}{~{\it et~al.}\ }
\newcommand{\eg}{{\it e.g.}}
\newcommand{\sfir}{S_{\rm FIR}}
\newcommand{\smin}{$S_{\rm min}$}
\newcommand{\smax}{$S_{\rm max}$}

\begin{document}

\title{Radio Wavelength Constraints on the Sources of the 
	\\ Far Infrared Background}

\author{D. B. Haarsma\altaffilmark{1} and R. B. Partridge\altaffilmark{2}}
\affil{Haverford College, Haverford, PA 19041}
\altaffiltext{1}{dhaarsma@haverford.edu}
\altaffiltext{2}{bpartrid@haverford.edu}


\begin{abstract}

The cosmic far infrared background detected recently by the COBE-DIRBE
team is presumably due, in large part, to the far infrared (FIR)
emission from all galaxies.  We take the well-established correlation
between FIR and radio luminosity for individual galaxies and apply it
to the FIR background.  We find that these sources make up about half
of the extragalactic radio background, the other half being due to
AGN.  This is in agreement with other radio observations, which leads
us to conclude that the FIR-radio correlation holds well for the very
faint sources making up the FIR background, and that the FIR
background is indeed due to star-formation activity (not AGN or other
possible sources).  If these star-forming galaxies have a radio
spectral index between 0.4 and 0.8, and make up 40 to 60\% of the
extragalactic radio background, we find that they have redshifts
between roughly 1 and 2, in agreement with recent estimates by
Madau\etal of the redshift of peak star-formation activity.  We
compare the observed extragalactic radio background to the integral
over the $\log N-\log S$ curve for star-forming radio sources, and
find that the slope of the curve must change significantly below about
1\mujy.  At 1\mujy, the faint radio source counts predict about 25
sources per square arcminute, and these will cause SIRTF to be confusion
limited at 160\mum.

\end{abstract}


\section{Introduction}

The COBE-DIRBE team has recently announced the detection of the cosmic
far infrared background (\cite{hauser98a}). The team interprets this
flux as the sum of far infrared (FIR) emission from all galaxies,
presumably mostly reemission from warm dust (although other
interpretations have been made, \eg\ \cite{bond86a}, 1991).
The DIRBE results thus provide important
constraints on the star formation history.  In this letter, we use the
DIRBE flux measurement, in conjunction with radio wavelength
observations, to explain some of the radio extragalactic surface
brightness, to estimate the redshift of the sources producing the
DIRBE background, and to constrain radio source counts.

At 240\mum\ and 140\mum, the fluxes measured by the DIRBE team are
$14\pm3\times10^{-9}\unit{Wm^{-2}ster^{-1}}$ and
$25\pm7\times10^{-9}\unit{Wm^{-2}ster^{-1}}$, respectively
(\cite{hauser98a}).  These values are in agreement with the FIRAS
spectrum over 125--2000\mum\ reported by Fixsen\etal(1998).
We have elected to use these more recent results rather than the
tentative detection reported by Puget\etal(1996).
The Hauser\etal measurement, and the
upper limits at 60\mum\ and 100\mum, show that the flux density has a
nearly flat spectrum, and thus we take the flux density to be
$\sfir=1.15\pm0.20\times10^{-20}$\wmsh\ between 140\mum\ and 240\mum\
in our subsequent calculations.

\section{Prediction of the Extragalactic Radio Background}

We begin with the assumption that the well-established correlation
between the far infrared and 20\cm\ radio flux densities of
galaxies,
\begin{equation}
	\frac{S_{80\mum} }{ S_{20\cm}} = 10^{2.34},
\end{equation}
(\cite{helou85a}; \cite{condon91a}) holds for whatever sources produce
the bulk of DIRBE background.  The relationship has been observed for
galaxies with a range of $10^4$ in FIR flux density, from normal
spirals to ultraluminous IRAS galaxies (\cite{cox88a};
\cite{crawford96a}).  The correlation does not, however, hold for
galaxies in which the radio emission is not associated with star
formation, such as AGN or classic radio galaxies.  The physical origin
of the correlation, while not completely understood, is probably due
to star-formation processes, namely dust reemission in the far
infrared, and supernova remnant synchrotron emission or thermal
emission in the radio (see \cite{condon92a} for a review).  Thus, it
is reasonable to suppose that the relationship holds for the higher
redshift, faint galaxies producing the DIRBE background.

A substantial fraction of the DIRBE background, and the associated
radio emission, may be redshifted.  Restframe emission at 80\mum\ from
sources in the range $0.75<z<2.0$ would be detected by DIRBE between
140\mum\ and 240\mum.  Since the luminosity density and star formation
rate peak approximately in this same interval (\cite{madau96a}), we
will initially take $z=1.0$ as a typical redshift for the sources
producing the DIRBE background.  Thus the radio flux emitted by these
sources at 20\cm\ (1.4~GHz) will be observed at 40\cm\ (750~MHz).

Using the FIR-radio correlation, we calculate that the surface
brightness at 40\cm\ due to star-forming galaxies is $S^\ast_{40\cm} =
10^{-2.34} \sfir \simeq 5300$~Jy/ster, or about 0.5~mJy/amin$^2$. The
asterisk is used to indicate emission from star-forming galaxies
(i.e., emission from AGN and the CMB is not included). The
corresponding brightness temperature is $T^{\ast}_{40\cm}= 0.31$~K.

To compare this to the observed brightness temperature, we scale this
result to other radio bands using an appropriate spectral index
$\alpha$ (where $S_\nu \propto\nu^{-\alpha}$) for the source
population.  Windhorst\etal(1993)
compiled various surveys of faint radio sources at 50\cm\ (600~MHz)
and 75\cm\ (400~MHz), and found the median spectral index between
these bands and 6\cm\ (5~GHz) to be $\alpha\sim0.7$ for faint sources,
a value we will assume unless noted.  Thus, the brightness temperature
at 170\cm\ (178~MHz) due to the DIRBE sources is
$T^\ast_{170\cm}\sim15$~K.

We may compare this prediction with the observations of Bridle (1967),
who found $T_{170\cm}=30\pm7$~K for the entire
extragalactic brightness temperature, including emission from AGN and
2.728~K from the cosmic microwave background (\cite{fixsen96a}).
Subtracting the temperature of the CMB, we see that source of the
DIRBE background explains more than half of the observed radio sky
brightness due to all extragalactic sources.  If we adopt a flatter
spectral index for the radio sources of $\alpha=0.4$, the fraction of
the observed radio background explained by the DIRBE sources drops to
about a third.  Thus the simple assumption that the FIR-radio
correlation holds for the DIRBE sources allows us to explain a
significant fraction of the observed radio brightness of the sky.

Our results are consistent with observations at other radio bands,
which show that about half of the extragalactic radio background is
due to star-forming galaxies.  For instance, Condon~(1989)
estimates that at 20\cm, AGN and star-forming galaxies contribute
almost equally to the total spectral power density.

An analysis of radio source counts allows us to reinforce this
conclusion.  The counts are the sum of the two populations
(star-forming galaxies and AGN), with the bulk of the AGN brighter
than the bulk of the star-forming galaxies (see, \eg,
\cite{windhorst93a}).  We can thus take all the bright sources as a
proxy for AGN, and then estimate the total fraction of the surface
brightness of the radio sky that AGN would contribute.  At 20\cm, the
dividing line between the two populations has been estimated as 1~mJy
(\cite{condon89a}) and 9~mJy (\cite{kron85a}).  Ryle~(1968)
uses a lower limit of 10~mJy at 75\cm\
(which corresponds to 4~mJy at 20\cm\ for $\alpha=0.7$), and 
integrates over $dN/dS$ to find $T^{{}^{\rm \!A\!G\!N}}_{75\cm} \sim
1.4$~K for the contribution to the background from AGN.  We can
compare this to the total radio background by scaling Bridle's 1967
measurement at 170\cm\ and correcting for the CMB
contribution to find $T_{75\cm}=2.9\pm 0.7$~K for the total radio
brightness.  Thus, the AGN explain about half of the total radio
background, leaving the other half due to star-forming systems.

Since observations at other radio bands have also found that about
half of the extragalactic radio background is produced by star-forming
galaxies, we conclude that the assumed FIR-radio correlation holds for
the sources responsible for the DIRBE background. Since the tight
FIR-radio correlation seen in individually observed galaxies at lower
redshift is ascribed to star formation, our results bolster the
argument that star formation is the cause of most of the FIR 
background at the presumably higher redshifts of the DIRBE sources.

\section{The Redshift of Sources Producing the DIRBE Background}

We have so far assumed that both the FIR and the radio flux are
produced at $z = 1$.  If the bulk of the emission originates at a
different redshift, we will need to make the appropriate
K-corrections.  It is convenient that the DIRBE measurements and
limits have found the FIR background to have a flat spectrum between
140 and 240\mum\ (see \S1), so that to first order we can ignore the FIR
K-corrections in the range $0.75<z<2.0$ for 80\mum\ emission.  At
radio wavelengths, the spectral index must be used to make the
K-corrections.

To find the redshift of the DIRBE sources, we once again assume the
FIR-radio correlation. We then scale the radio emission to an
observing wavelength of 170\cm, and write $A$ as the fraction
$S^\ast_{170\cm} / S_{170\cm}$ of the observed flux density which we
ascribe to star-forming galaxies, giving
\begin{equation}
	S_{170\cm} = \frac{1}{A} \sfir 10^{-2.34}
			\left( \frac{170\cm}{20\cm(1+z)} \right)^{\alpha}.
\end{equation}
Solving this expression for the redshift, and substituting in the
measured values of $T_{170\cm}$ (see \S2) and $\sfir$ (see \S1), we
find the dimensionless relationship
\begin{equation}
	A \left( \frac{1+z}{8.5} \right) ^\alpha = 0.20 \pm 0.06,
\label{eq.Azalpha}
\end{equation}
which is plotted in Figure~1.  

For our nominal values of $A=0.5$ and $\alpha=0.7$, we find
$z+1\sim2.3\pm1.0$, in agreement with our initial assumption of $z=1$.
While the uncertainty is large, this value is in agreement with the
redshift of peak star formation of $z\sim1.5$ found by Madau\etal(1996).
The measured redshifts of individual faint radio galaxies (less than
1~mJy) are typically $z\sim0.5-0.75$ (\cite{condon89a};
\cite{windhorst93a}; \cite{richards98a}); our value of $z$ is somewhat
higher than this estimate, but this is reasonable since the fainter
galaxies making up the DIRBE background should be at higher redshifts
than brighter galaxies detected individually.

Let us now assume reasonable values of $z$ and $\alpha$ and consider
the allowable range of the parameter $A$.  We can assume that the bulk
of star formation is certainly more recent than the epoch
corresponding to $z = 8$.  Then Figure~1 shows that the fraction of
the radio background from star forming galaxies, $A$, must be greater
than 20\% for any reasonable spectral index.  Since $z$ must be
greater than zero, we can also find upper limits on $A$: for
$\alpha=0.7$, $A$ must be less than about 90\%, and for $\alpha=0.4$,
less than 50\%.  If star formation peaks in the interval $1<z<2$ and
$\alpha\sim0.7$, $A$ is tightly constrained to the interval 40-60\%.

\section{Constraints on Faint Radio Source Counts}

The brightness temperature of the radio background, corrected for the
CMB, is due to combined flux of the faint radio galaxies and thus is
related to the number counts of these galaxies,
\begin{equation}
	T = \frac{\lambda^2}{2k} \int_{S_{\rm min}}^{S_{\rm max}} 
			S \frac{dN}{dS} dS.
\end{equation}
We can take the number counts to be of the form $dN/dS = C S^\gamma$,
where $\gamma$ is generally less than $-2$ for this population at
radio wavelengths (see eq.~[\ref{eq.dnds}]).  For $\gamma<-2$, faint
sources dominate the contribution to the background temperature. It
has also long been recognized that, for $\gamma<-2$, the total source
counts and the radio sky brightness will (slowly) diverge as $S
\rightarrow 0$ (see, \eg, \cite{windhorst93a}).  Therefore, there must
be a cutoff or change in slope of the $\log N-\log S$ curve at some
minimum value, \smin.  To estimate \smin\ we can use the radio
brightness temperature due to the DIRBE star-forming galaxies in
conjunction with the $\log N - \log S$ relation for faint radio
sources.

Very deep VLA observations at 3.6\cm\ (8~GHz) (\cite{windhorst93a};
\cite{windhorst95a}; \cite{fomalont97a}; \cite{kellermann98a})
have been used to determine the number count power law,
\begin{equation}
	\frac{dN}{dS} = - 4.6\pm0.7
		\left( \frac{S}{1\unit{Jy}} \right)^{-2.3\pm 0.2}
		\unit{Jy^{-1}ster^{-1}},
\label{eq.dnds}
\end{equation}
which is valid for flux densities in the range $14.5\mujy <S <
1000\mujy$.  At these faint flux levels, there is little contamination
from AGN, i.e. the sources are nearly all star forming galaxies
(\cite{condon89a}).  For $\gamma<-2$, the definite integral is
dominated by \smin\ and is independent of \smax\ for large values of
\smax; we take \smax\ to be infinity.  Thus, the integral over these
number counts will yield the radio brightness temperature due to the DIRBE
sources found above.  Scaling this temperature to 3.6\cm, integrating,
and solving for \smin, we find
\begin{equation}
	S_{min} = \left[ \frac{2+\gamma}{C} 
			\left( \frac{\sfir}{\rm Jy} \right)
			 10^{-2.34}
			\left( \frac{3.6\cm}{20\cm(1+z)} 
				\right)^\alpha
		\right]^{\frac{1}{2+\gamma}}.
\label{eq.smin}
\end{equation}
For $\alpha=0.7$, $z=1.0$, the observed value of $\sfir$, and $\gamma$
and $C$ from equation~(\ref{eq.dnds}), we find $S_{min} \gtrsim
1\mujy$ at 3.6\cm.  

It is useful to compare our estimate of \smin, based on the far
infrared background, with other estimates at 3.6\cm, such as those
made by Windhorst\etal(1993).
In order for the radio background due to galaxies to not  distort the
spectrum of the cosmic microwave background (CMB), they found $S_{min}
\geq 20$~nJy at 3.6\cm.  They also found that if \smin\ fell below
300~nJy, the optical counterparts of these faint radio sources would
exceed the V band counts of field galaxies.  The limit of $S_{min}
\geq 1$\mujy\ we have found is consistent with these, but more
restrictive, and has more interesting observational consequences.

An RMS sensitivity of 1.5\mujy\ has been reached in recent VLA
observations at 3.6\cm\ (\cite{partridge97a}), and our value of \smin\
indicates that there are few radio sources below this flux density,
that is, the slope of the $\log N-\log S$ curve changes significantly
at \smin.  The present VLA observations have thus detected the bulk of
all radio sources in the Universe.

The value of \smin\ (at $z\sim1$) can also be compared to the knee in the
luminosity function for {\em local} faint radio sources.  At 20\cm,
the slope turns over around $10^{22.4}$~W/Hz (\cite{condon89a}).  A
source at $z=1$ with a flux density of 1\mujy\ at 3.6\cm\ would have a
20\cm\ luminosity of about $10^{21.7}h^{-2}$~W/Hz (assuming
$\alpha=0.7$ and an Einstein-deSitter universe).  Thus, the turnover
point at $z=1$ is very similar to the local value, especially for a
Hubble parameter of $h\sim0.6$.

We can also consider the fluctuations in the radio background due to
these sources.  The number counts (eq.~[\ref{eq.dnds}]) at the 1\mujy\
level indicate that there should be about 25 sources per armin$^2$, or
about $3\times10^8$~sources/steradian.  The average separation of
1\mujy\ sources is thus about 14$''$, and we would expect substantial
fluctuation in the surface brightness of the radio sky on this angular
scale.  The RMS due to the sources can in principle be calculated from
the number counts, assuming the sources are distributed randomly on
the sky.  When the number counts follow a power law distribution, as
in equation~(\ref{eq.dnds}), the RMS is determined by the upper and lower
limits on the source flux densities, \smin\ and \smax.  For a power
law with $\gamma>-3$, the RMS is unfortunately dominated by the value
of \smax, which is not well known for this population (the bright
star-forming galaxies are comparable to the faint AGN, making it
difficult to determine the precise upper bound on the population).
Thus, it goes beyond the scope of this paper to estimate the RMS in
the radio background due to this population.  We concur, however, with
the conclusions of Partridge\etal(1997)
and Mitchell \&~Condon~(1985),
that these faint radio
sources do contribute substantially to the RMS fluctuations in the
microwave sky. Others have noted that at the higher frequencies and
lower angular resolutions planned for satellite- and ground-based
searches for CMB fluctuations, the faint sources that produce the FIR
background should present no problems (\cite{windhorst93a};
\cite{toffolatti95a}), unless the sources are strongly clustered.

We can also ask how these sources will appear in number counts and
high resolution maps in the far infrared.  For $\alpha$ between 0.4
and 0.7, and $z=1$, we use the FIR-radio correlation to calculate that
3.6\cm\ sources with flux density \smin\ (1\mujy) will appear as
600-1200\mujy\ sources at 160\mum.  The number density of these faint
FIR sources can be compared to the semi-analytic models of galaxy
formation of Guiderdoni\etal~(1998),
who predict the FIR background and the faint galaxy counts for various
FIR wavelengths.  One of their models finds a number density of
10$^8$~sources/steradian for 1~mJy sources at 175\mum, which is close
to our estimate of $3\times10^8$ sources/steradian.  At 160\mum, the
longest SIRTF-MIPS wavelength, SIRTF will have 15$''$ pixels and an
angular resolution of about 30$''$ (\cite{rieke96a}), larger than the
average separation of 14$''$ between the faintest sources.  Thus,
SIRTF will not be able to resolve the individual sources responsible
for the DIRBE flux, and will be confusion limited.  This is in
reasonable agreement with the Monte Carlo simulations of
Rieke\etal(1996),
who find a confusion limit for SIRTF of 1500\mujy.

\section{Conclusions}

Using the well-established correlation between the FIR and radio
luminosity of individual galaxies, we have extrapolated the FIR
background detected by DIRBE to radio wavelengths.  This has allowed
us to derive several properties of the sources making up the
FIR background:
\begin{enumerate}

\item
The radio emission from the sources makes up about half of the
observed extragalactic radio background (about 0.3~K at wavelengths
around 40\cm, excluding the cosmic microwave background).

\item
Since this finding is in agreement with other radio observations, the
FIR-radio correlation holds even for the very faint sources making up
the DIRBE background.  This implies that the FIR background between
about 140 and 240\mum\ is dominated by star-formation, not AGN
activity.

\item
The typical redshift and spectral index $\alpha$ of these sources, and
the contribution they make to the radio background are related by
equation~(\ref{eq.Azalpha}).  For reasonable values of the spectral index
(0.4 to 0.7) and the fraction of the radio background (40-60\%), we
find the redshift of these sources to be roughly between 1 and 2, in
agreement with the Madau\etal(1996)
estimate of the redshift of peak star formation.

\item
By extrapolating the radio $\log N-\log S$ curve at 3.6\cm\ to fainter
flux densities, we estimate that most of the DIRBE flux is produced by
sources whose 3.6\cm\ flux density is $\gtrsim 1$\mujy, and that the
number density of 1\mujy\ sources is about $25/\unit{arcmin^2}$.
These sources will cause SIRTF to be confusion limited around 160\mum.

\end{enumerate}

\acknowledgements

We thank Steve Boughn, Jim Condon, Alberto Franceschini, Michael
Hauser, Ken Kellermann, Rogier Windhorst, and our referee for their
helpful comments.  This work was supported by NSF grant AST96-16971.

\clearpage



\clearpage
 
\begin{figure}
\plotone{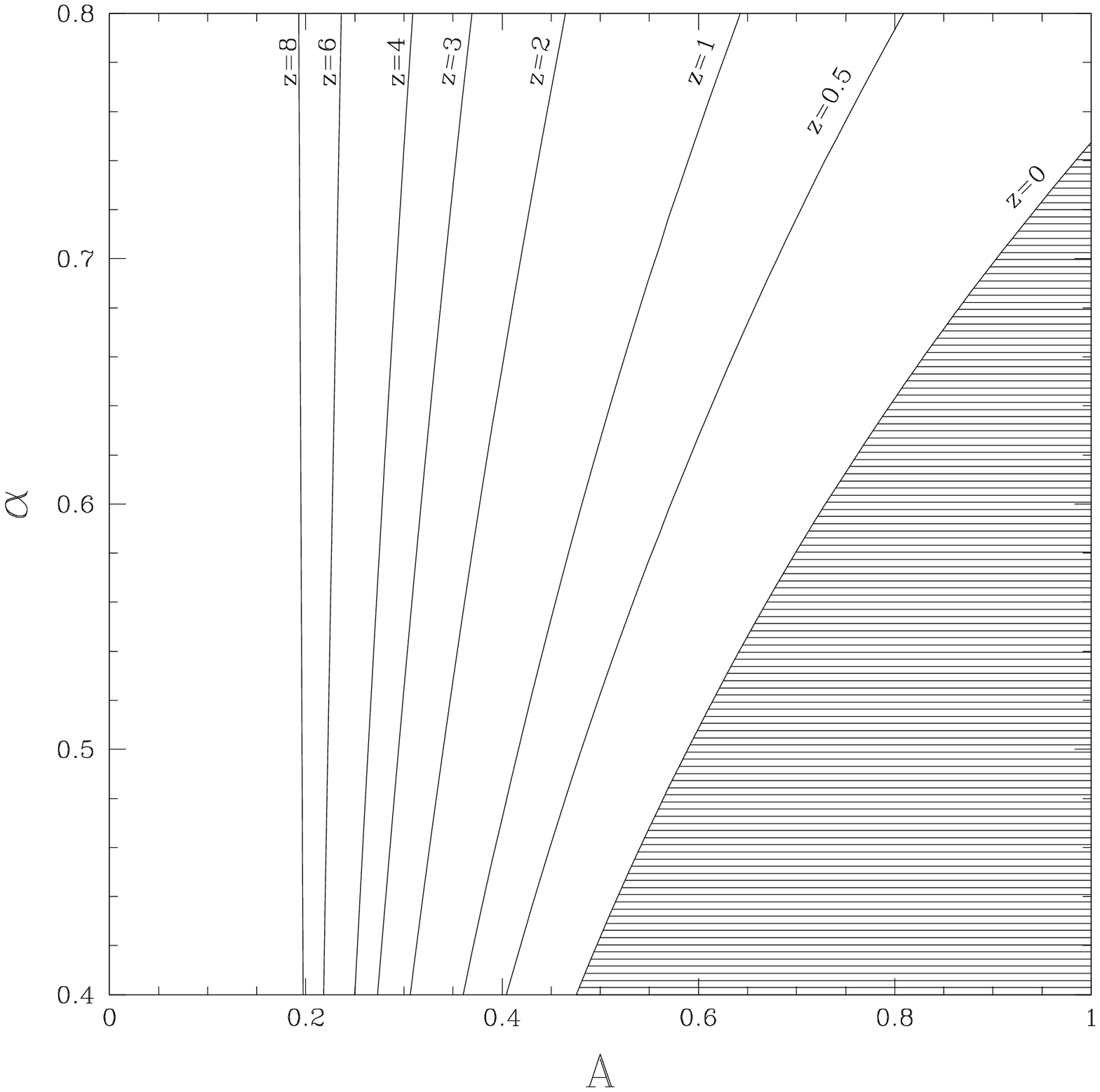}
\caption{
Relationship between the radio spectral index $\alpha$, the
ratio of star-formation flux to the total radio background $A$, and
the typical redshift $z$ for the sources making up the FIR
background.  See equation~(\ref{eq.Azalpha}).  
}
\end{figure}

\end{document}